\documentclass[11pt]{article}
\usepackage{amsmath}
\usepackage{graphicx} 
\usepackage{amsfonts}
\usepackage{amssymb}
\usepackage{epsf}
\usepackage{latexsym}
\newcommand{\be}{\begin{equation}}
\newcommand{\ee}{\end{equation}}
\newcommand{\bea}{\begin{eqnarray}}
\newcommand{\eea}{\end{eqnarray}}
\newcommand{\pa}{\partial}
\newcommand{\fn}{\footnote}

\begin{document}
\begin{titlepage}



\begin{center}
\Large
{\bf Geodesics, the Equivalence Principle and Singularities \\
in Higher-dimensional General Relativity and Braneworlds}
\vspace{.3in}
\normalsize

\large{Edward Anderson$^{1, 2, \dagger}$ and Reza Tavakol$^{2,
\ddagger}$}

\normalsize
\vspace{.3in}

{\em $^1$ 
Department of Physics, P-412 Avadh Bhatia Physics 
Laboratory, University of Alberta, 

Edmonton, Alberta, Canada T6G 2J1;

Peterhouse ,Cambridge U.K. CB2 1RD;

DAMTP, Centre for Mathemetical Sciences, Wilberforce Road Cambridge U.K. CB3 0WA.}

\vspace{.15in}

{\em $^2$  Astronomy Unit, School of Mathematical Sciences, Queen Mary,

University of London, U.K. E1 4NS}

\end{center}

\vspace{.3in}

\begin{abstract}

The geodesics of a spacetime seldom coincide with those of an 
embedded submanifold of codimension one. 
We investigate this issue for higher-dimensional general 
relativity-like models, firstly in the simpler case
without branes to isolate which features are already
present, and then in the more complicated case with branes. 
The framework in which we consider branes is
general enough to include asymmetric braneworlds
but not thick branes.  
We apply our results on geodesics
to study both the equivalence principle and cosmological
singularities. 
Among the models we study
these considerations favour $Z_2$ symmetric braneworlds
with a negative bulk cosmological constant.

\end{abstract}

PACS NUMBERS: 04.50.+h, 98.80.Jk

\vspace{.7in}
Electronic mail: $^{\dagger} $ea212@cam.ac.uk,
$^{\ddagger}$R.Tavakol@qmul.ac.uk

\end{titlepage}


\setcounter{equation}{0}

\section{Introduction}

Higher dimensional general relativity (GR) theories have recently
attracted much attention. The main role of the extra dimension 
in such
theories is to provide extra degrees of freedom on the $3+1$ 
hypersurface
which is taken to represent the apparent world.  Although these 
are not
leading candidates for fundamental theories, they have been used to 
study some
aspects of string/M-theory.

Examples of such higher-d GR theories include (noncompactified)
$(n + 1)$-d GR \cite{ndGR} (where the choice of the 3 + 1 apparent world
is arbitrary), Kaluza--Klein theory \cite{KK} (where
one dimension is compactified leading to a $3 + 1$ apparent world described
by GR, electromagnetism and a scalar), low energy string cosmology
\cite{stringcos} (based on generalized compactifications), and
some of the simpler braneworlds \cite{branes, SMS, Maartensrev, BC,
asymbranes} (in which the apparent world consists of matter and tension
restricted to a special {\it brane} hypersurface and the actual world 
is this hypersurface together with a surrounding {\it bulk} spacetime).

Given that the observed macroscopic apparent world is $(3 + 1)$-d,
an important question concerning such theories is
how closely their corresponding
$3 + 1$ apparent worlds resemble the observed world, and in
particular how observations made in their
apparent worlds can test and set constraints on
the full theories themselves.

The first aim of this article is to clarify the extent to which
such theories satisfy geodesic postulates similar to those of 4-d GR,
concerning the motion of particles and light rays.
While works relating the geodesics of an embedded 3 + 1 apparent world 
spacetime and those of the embedding 4 + 1 spacetime
have recently appeared (see e.g \cite{Ishihara,Seahra}), we
generalize these and work out further consequences of this relation, 
for 
5-d GR models in Sec. 2 and for a wide range of 5-d GR braneworld 
models
in Sec. 3.
We find a number of conditions under which the 3 + 1 and 4 + 1
geodesics coincide on the apparent world.  These include the so-called
$Z_2$ symmetry as one of the most robust conditions that
enforces this important coincidence.
This provides an important extra {\sl physical}
motivation for the consideration of $Z_2$ symmetric brane models
at the level of GR, independently of the usual motivations based on
heterotic
M-theory \cite{Davis, Maartensrev} and mathematical simplicity
\cite{Davis}.

Geodesics are central to the study of singularities.
Having studied the geodesics in higher-d GR models,
the second aim of this article is to consider singularities (and 
whether there
is any hope of their resolution) in these settings.
We shall study this question for higher-d GR (Sec. 4), 
as well as for a range of higher-d GR braneworld models (Sec. 5).
We shall see that such models become especially interesting in the 
latter case, with both $Z_2$ and non-$Z_2$ models giving rise to
a range of new possibilities.  We conclude in Sec. 6. 

\section{Foundations for GR and higher-d GR theories}

In this section we shall study the extent to which higher-d GR theories
are built on foundations analogous to those of GR. We recall
that the conventional formulation of GR has
three main postulates. (i) 3 + 1 spacetime is a pseudo-Riemannian
manifold, which is assumed to be the arena in which real
physical events and interactions take place,
(ii) Einstein field equations hold and 
(iii) light rays and freely-falling non-rotating, test-mass particles
respectively follow the null and timelike geodesics 
of the spacetime\footnote{We note that there is a long-standing
and ongoing debate within GR concerning the autonomy of this 
assumption, i.e.
whether these are additional postulates or whether the geodesic 
equations are
consequences of the Einstein field equations (see e.g \cite{Ehlers}).}. This allows contact 
to be
made with observations, and is intimately connected with the 
equivalence principle, according to which freely-falling non-rotating test
particles follow timelike geodesics irrespective of their composition (see \cite{EP, Will} 
for a thorough discussion).

\mbox{ }

We note that assumptions made in developing higher-d theories can be
viewed as more elaborate versions of the above.  This involves defining
the roles given both to the hypothesized {\it actual} $n$ + 1 world
and to the 3 + 1 {\it apparent} world. Concerning the field equations, 
the usual and simplest
practice\fn{One could also include other terms expected in a
low-energy effective action, such as the second-order Lovelock terms
\cite{Lovelock} that are nonzero for higher-d theories with
$n + 1 > 4$.  Including these gives a better approximation than GR to
string theory in certain regimes \cite{GrossSloan}. These terms are
included in some
braneworld studies \cite{GBbranes}.}
is to take $n = 4$ and employ the 4 + 1 Einstein field equations. While this would provide
a simple choice for the 4 + 1 field equations, it would cause the 3 + 1 {\sl apparent-world} 
field equations to differ from the 3 + 1 GR Einstein field equations, as a result of projections of
higher-d matter and `matter' induced terms from the higher-d geometry
\cite{ndGR,SMS}.

A crucial issue in braneworld/higher-d models
concerns the precise nature of the bulk/higher
dimensions. This can either be specified
by a fundamental theory, in which case it would in general constrain
the dynamics of the brane/lower-d apparent world with observational consequences;
or it can be left free - as is to some extent true
in most braneworld models considered so far - in which case the lower-d
physics will remain unrestricted, unable to
provide constraints which may be compared with observations.
Clearly the former is essential for  physically complete
models. In the following we shall mainly consider vacuum or Einstein
space bulks, which cover a range of scenarios \cite{ndGR, branes, 
SMS}.

Finally, one would require the analogues of the geodesic postulates
in GR, both in order to make contact with observations, and because
these are implicit in the set-up of some higher-d scenarios where
some fields are assumed to be 4-d while others are allowed to propagate
in 5-d.  Geodesics are also crucial in discussing singularities for 
these
higher-d GR models.  In this section we consider 4 + 1 GR including the
assumption that the 4 + 1 version of the equivalence principle holds, in the sense that 
freely
falling non-rotating test-mass particles follow 4 + 1 geodesics 
irrespective
of their composition.
Now since all test masses follow the same paths in (4 + 1)-d,
they will also appear to follow the same paths from the perspective
of any included 3 + 1 apparent worlds.  Thus there will be some kind
of `equivalence principle' from the apparent world perspective.  However, the apparent world
will not in general look like a 3 + 1 GR world because the set
of curves privileged by free fall in the apparent world will {\sl not}
in general include the 3 + 1 timelike geodesics privileged in 3 + 1 GR.
Likewise the apparent world's set of curves privileged by light rays
will {\sl not} in general include the null geodesics privileged in 3 + 1 GR.  This
is made clear by the following comparative study of the geodesic 
equations
in each space, which is more general that previous studies
\cite{Israel,Ishihara}, since we allow general rather than merely
tangential paths from the apparent-world perspective.

\mbox{ }

Contrast the geodesic equations\fn{For ordinary GR, 
4 + 1 bulk and 3 + 1 apparent world spacetimes we use capital 
sans serif, bold and plain indices respectively.  
For spatial slices of each of these we use the corresponding 
lower-case indices.  Note that $t$ and $z$ indices are reserved 
to indicate the timelike direction and the 
perpendicular direction away from the apparent world.   
$\mbox{\sffamily x}^{\mbox{\scriptsize{\sffamily A}}}$, 
$\mbox{\bf x}^{\mbox{\bf{\scriptsize A}}}$ and $x^A$ 
are coordinates for ordinary GR, the 4 + 1 bulk and 3 + 1 
apparent worlds respectively.  
$\mbox{\sffamily D}_{\mbox{\scriptsize{\sffamily A}}}$, 
$\mbox{\bf D}_{\mbox{\bf{\scriptsize A}}}$ and $D_A$ 
are covariant derivatives for each of these.  
The dot denotes time derivative.} of 3 + 1 GR 
\be
0 = \dot{\mbox{\sffamily x}}^{\mbox{\scriptsize{\sffamily A}}}
\mbox{\sffamily D}_{\mbox{\scriptsize{\sffamily A}}}
\dot{\mbox{\sffamily x}}^{\mbox{\scriptsize{\sffamily B}}}
\ee
with the geodesic equations of a 4 + 1 bulk \fn{It should be noted 
that a second connection, namely the 4 + 1 metric connection, comes 
into physics via the equation of motion of test particles. This 
raises the question of whether from the apparent-world perspective one has a 
metric-affine theory rather than a GR theory and whether the 
connection being metric suffices in proposed braneworld set-ups.  
These foundational issues are equally relevant when branes are introduced.}
\be
\label{4-1geod} 
0 = \dot{\mbox{\bf x}}^{\mbox{\bf{\scriptsize A}}}
\mbox{\bf D}_{\mbox{\bf{\scriptsize A}}}
\dot{\mbox{\bf x}}^{\mbox{\bf{\scriptsize B}}}.
\ee
Splitting these equations with respect to a candidate
3 + 1 apparent-world hypersurface we have
\be
0 =
\left\{
\begin{array}{c}
\dot{x}^A \mbox{\bf D}_A \dot{x}^B + \dot{x}^{z} \mbox{\bf
D}_{z}
\dot{x}^B \\
\dot{x}^A \mbox{\bf D}_A \dot{x}^{z} + \dot{x}^{z} \mbox{\bf
D}_{z}
\dot{x}^{z}
\end{array}
\right\} \mbox{ } ,
\ee
where z is the bulk coordinate, and where we have used the 
splitting with respect to a $(n - 1) + 1$ hypersurface for a 
$(1, 0)$ tensor on an $n$ + 1 space,
$\dot{\mbox{\bf{x}}}^{\mbox{\scriptsize{\bf A}}} = \{\dot{x}^A,
\dot{x}^{z}\}$.   
Using the corresponding splitting for the covariant derivative
$\mbox{\bf D}^A$ as can be simply deduced from e.g 
\cite{KucharII} and passing to normal coordinates, 
the geodesic equation (\ref{4-1geod}) becomes 
\be
0 =
\left\{
\begin{array}{c}
\dot{x}^AD_A\dot{x}^B - 2\dot{x}^A {K_A}^B\dot{x}^{z} +
\dot{x}^{z}\frac{\pa \dot{x}^{B}}{\pa z}\\
\dot{x}^AD_A\dot{x}^{z} + K_{AB}\dot{x}^A\dot{x}^B +
\dot{x}^{z}\frac{\pa \dot{x}^{z}}{\pa z}
\end{array}
\right\} \mbox{ } .
\ee
One can view these as representing the motion under velocity-dependent
force terms, including terms similar to the Lorentz force, with $x^{z}$
playing the role of charge and the extrinsic curvature $K_{AB}$ 
playing that of the electromagnetic field tensor $F_{AB}$.  
We note that for motions with $x^{z} \neq 0$
and assuming $K_{AB} \neq 0$, one would obtain different
5-d trajectories for different $x^{z}$, similar to the way
particles with different charges travel on different paths
in nonzero electromagnetic fields.

For the geodesics to coincide, we shall allow an arbitrary matter flow
$\dot{\mbox{\bf{x}}}^{\mbox{\bf{\scriptsize A}}}$ and then ask
under what conditions would  the candidate 3 + 1 apparent world
strictly reproduce GR, by demanding that
\be
\dot{\mbox{\bf x}}^{\mbox{\bf{\scriptsize A}}}
\mbox{\bf D}_{\mbox{\bf{\scriptsize A}}}
\dot{\mbox{\bf x}}^{\mbox{\bf{\scriptsize B}}} =
\left\{
\begin{array}{c}
\dot{x}^AD_A\dot{x}^B \\ 0
\end{array}
\right\}
\mbox{ } .
\ee
One has coincidence if \be
\dot{x}_{z} = 0 ~~\mbox{  and }~~  K_{AB} = 0
\mbox{  } ,
\ee
\be
\mbox{\sl{or} if } \mbox{ } \mbox{ }
    \frac{      \pa\dot{\mbox{x}}^{A}    }{\pa z} = 0 ~~
\mbox{ } \dot{x}^{z} = \mbox{ constant } ~~ \mbox{ and }~~ K_{AB} = 0
\mbox{ } .
\ee
These conditions ensure {\sl both} that the 3 + 1 equivalence principle holds
and that no matter leaves the 3 + 1 apparent world.

The condition $K_{AB}=0$, which is crucial in both cases,
is termed the
{\sl totally
geodesic
condition}\fn{This geometrical significance of
$K_{AB} = 0$ has been known at least since Eisenhart \cite{KABzero}.}
for
a
3 + 1
hypersurface embedded within a 4 + 1 bulk.
The significance of $K_{AB} = 0$
has already been noted in the higher-d literature
(e.g \cite{Seahra}), for the cases with $\dot{x}^{z}=0$.
The advantages of not presupposing this condition are twofold.
Firstly it leads to the second case above,
that is an example of a 3 + 1 world hypersurface
which, while totally geodesic, allows flows
in the perpendicular direction at the same rate.
Furthermore there is no difference between the on-apparent-world
components of the flows and the corresponding components of
flows on a nearby `parallel' 3 + 1 world.  
Secondly this highlights the fact that leaking matter and equivalence principle violation
effects are two sides of the same coin in setting up apparent worlds
within higher-d GR bulks.

While the assumption of total geodesy
would ensure that both astronomical observations and equivalence principle experiments
are in accord with accepted  physics and that no matter
would disappear from the 3 + 1 apparent world, it is important to note that 
{\sl 3 + 1 apparent worlds cannot generically be accommodated as
totally-geodesic hypersurfaces within e.g 4 + 1 vacuum manifolds.}
This is due to the fact that the specification of a general apparent world 
would require full knowledge of the form of its metric $h_{AB}$
(10 independent pieces of information), as well as the simultaneous
assumption of total geodesy (i.e. $K_{AB} = 0$, which amounts to 10
more pieces of information). But, as $h_{AB}$ and
$K_{AB}$ are restricted by 5 constraint equations,
this specification 
is 
in general impossible.
By the Gauss constraint, only those apparent worlds with Ricci scalar 
$R = 0$ can be treated thus.
One might attempt to avoid this difficulty by leaving the bulk matter 
content
unspecified, in which case the constraint equations could be viewed as
equations to solve for the bulk matter content that permits such a 
slice to
exist. Generically, however, there is no reason for the ensuing
bulk matter content to be physically reasonable, which is similar to 
the related problem in GR where arbitrary specification of the metric
on the left hand side of Einstein field equations does not generally result in a reasonable 
energy momentum tensor on the right hand side, 
as was pointed out by Synge \cite{Synge} long ago.  
As we shall see total geodesy is both restrictive and 
supplantable for $Z_2$ braneworlds (Sec. 3).

\mbox{  }  

One could try to remedy this problem by considering `almost total 
geodesy',
instead, where the `almost' implies `within observational bounds'.
We recall that observations of bending of starlight by the the Sun and 
galaxies
(by microlensing), together with the observations of the Shapiro effect
conform to geodesic motion. Furthermore, the universality of free fall 
is
known extremely accurately for a broad range of compositions.
Even though it is not these observations that are under direct
consideration here, clearly a substantial breakdown of the geodesic
assumption would compromise the interpretation of the above 
observations
within a GR framework.
On the other hand these are not clean tests of geodesic motion.
Indeed, the presupposition  of the geodesic principle
runs so deep in the GR literature that we were unable to find
how accurately it is known that light and freely-falling
particles follow geodesics. Therefore it is possible that perfect
and universal geodesic motion only hold approximately
were we to live on an apparent world, rather than a 4-d GR world.
In that case the fundamental theory should provide
convincing (i.e. not fine-tuned) arguments for why the departures from
perfect geodesic motion should be small, while the 
theoretical framework \cite{Will} of experimental relativity  
should provide accurate bounds on this smallness.

There is also the associated difficulty of what would be the
physical status of an apparent world, if it leaked matter, even slowly.
It might be argued that this problem would be resolved
by identifying the apparent world with the hypersurface
which `goes with the flow', rather than as a leaky hypersurface which
just happens to possess a theoretically-desirable metric.
However, two issues would in general still remain:
does the apparent world become distorted or crinkled if $K_{AB}$ is
inhomogeneous and could the flow
lead to `spreading/incoherence', in the sense that  particles of 
different
composition or moving at different 3 + 1 velocities could flow at 
different
rates within the 4 + 1 picture, so that an initial thin sheet could
become much thicker at later times. [We shall return to this question
in Sec. 3].

Finally there is the question of how to mathematically quantify the
notion of almost geodesy.  Comparison of curves only makes sense to us
if they are on the same manifold and if they share endpoints
[\cite{Wald} contains a method of doing this].  

\section{Do higher-d GR braneworlds have analogous foundations?}

In this section we extend the investigation of Sec. 2 to the
case of higher-d GR braneworld models in which the apparent world 
is confined to an infinitely thin brane which generally has both
tension and matter. The models considered here have the
important advantage over those in Sec. 2 that they can accommodate a
wide range of apparent worlds, thus making contact with a broad range of
braneworld models \cite{branes, SMS, Maartensrev, BC, asymbranes}.

Consider a general notion of braneworld with thin branes which do not
necessarily satisfy the $Z_2$ symmetry, i.e which have a distinct
bulk spacetime either side of the brane.  Braneworld scenarios usually
involve a more complicated set of physically privileged curves than in
GR.  In such models, it is usually postulated that gravitons and
bulk-species test matter move on 4 + 1 geodesics,
while ordinary test matter follows curves
that are confined to a 3 + 1 apparent world.  
Thus these postulates
can amount to a significant practical alteration of the foundations of
physics. Thus from a 4 + 1 GR perspective the usual notion of geodesics
is applied to the 4 + 1 ones, which is now reserved for the motion of
speculative bulk matter, while all brane-confined matter fields (test
particles, light rays) are postulated to move on 3 + 1 geodesics.
Since as we saw above these are not in general among the 4 + 1 
geodesics,
one is in effect postulating two separate privileged classes of curves
for free fall.  This, by definition, amounts to equivalence principle violation and in 
this
sense these models are not generally expected to be higher-d {\sl GR}
theories.  But unless there is little or no equivalence principle violation, they would
not be viable observationally.

We begin our analysis by looking for conditions which ensure that the
geodesics on the brane and bulk coincide. We shall use the notation
\be
[A] = A_+ - A_- \mbox{ } ,
\ee
\be
(A) = A_+ + A_- \mbox{ } , 
\ee
where +, -- denote each side of the brane.
For tractability we require a number of objects to be continuous across
the brane, namely the apparent world metric:
\be
[g_{AB}] = 0 \mbox{ } ,  \mbox{ }
\ee
and some velocities 
\be
[\dot{x}_A] = 0 \mbox{ } , \mbox{ } [\dot{x}_{z}] = 0 \mbox{ } .
\ee
Moreover a discontinuity is permitted in $K_{AB}$ in the
limit of a thin sheet of matter being sandwiched between the two bulk
portions so that \cite{Israel} 
\be
-\frac{1}{2}Y_{AB} = [{K}_{AB} - g_{AB}K]
\mbox{ } \mbox{ } \Leftrightarrow \mbox{ } \mbox{ }
[K_{AB}] = - \frac{1}{2}\left\{ Y_{AB} - \frac{Y}{3}g_{AB}\right\}
\label{new} \mbox{ } , 
\ee
for $K$ is the trace of the apparent world's  
extrinsic curvature $K_{AB}$ and 
$Y_{AB}$ is composed of the thin matter sheet energy-momentum $T_{AB}$ and the
brane's tension $\lambda$:
\be
Y_{AB} = T_{AB} - \lambda g_{AB} \mbox{ } .
\ee
The Einstein field equations for the composite spacetime is then 
$\mbox{\bf G}_{\mbox{\bf{\scriptsize AB}}} =
\mbox{\bf T}_{\mbox{\bf{\scriptsize AB}}} - \Lambda
\mbox{\bf g}_{\mbox{\bf{\scriptsize AB}}} -
\delta^{(4)}(z)Y_{\mbox{\bf{\scriptsize AB}}}$, 
where  $\mbox{\bf g}_{\mbox{\bf{\scriptsize AB}}}$, 
$\mbox{\bf G}_{\mbox{\bf{\scriptsize AB}}}$,  
$\mbox{\bf T}_{\mbox{\bf{\scriptsize AB}}}$ 
and $\Lambda$ are the bulk metric, Einstein tensor, energy--momentum and 
cosmological constant respectively.

As regards geodesics, what is relevant on-brane are the geodesic
equations with the {\sl average} bulk
contributions.\fn{In writing this down, we simplify by using $A$ in
place of $(A)/2$ for all the objects we have declared to be
continuous.} Thus one generally has, in `brane-centric' normal 
coordinates,
\be
0 =
\left\{
\begin{array}{c}
\dot{x}^AD_A\dot{x}^B - \dot{x}^A {(K_A}^B)\dot{x}^{z} +
\dot{x}^{z}\frac{\pa \dot{x}^{B}}{\pa z}\\
\dot{x}^AD_A\dot{x}^{z} + \frac{1}{2}(K_{AB})\dot{x}^A\dot{x}^B +
\dot{x}^{z}\frac{\pa \dot{x}^{z}}{\pa z}
\end{array}
\right\}
\mbox{ } .
\ee
The condition for the 3 + 1 and 4 + 1 geodesics to coincide now becomes
\be
0 = \dot{\mbox{\bf x}}^{\mbox{\bf{\scriptsize A}}}
\mbox{\bf D}_{\mbox{\bf{\scriptsize A}}}
\dot{\mbox{\bf x}}^{\mbox{\bf{\scriptsize B}}} =
\left\{
\begin{array}{c}
\dot{x}^AD_A\dot{x}^B \\ 0
\end{array}
\right\}
\mbox{ } .
\ee
which holds if
\be
\dot{x}^{z} = 0 ~~ \mbox{and} ~~(K_{AB}) = 0 \mbox{ } ,
\ee
\be
\mbox{ or if } \mbox{ } \mbox{ }
\frac{\pa\dot{\mbox{x}}^{\mbox{{\scriptsize A}}}}{\pa z} = 0 \mbox{ , }~~
\dot{x}^{z} = \mbox{ constant }
\mbox{ and } ~~
(K_{AB}) = 0
\mbox{ } .
\ee

Comparing with the previous section, we note that from (\ref{new}) 
total geodesy now
requires that the on-brane matter has a pure trace energy--momentum 
tensor.
Furthermore, it is now the {\sl averaged} totally-geodesic condition 
$(K_{AB}) = 0$
(also known as $Z_2$ symmetry) which ensures the coincidence of the 
apparent and bulk
geodesic equations (i.e. no equivalence principle violation) as well as ensuring that bulk 
matter does not fall on or
off the apparent world. While this condition does not generally
hold for braneworld scenarios, it does hold for a {\sl
range}\fn{This permits apparent worlds containing {\sl minimally coupled} scalar,
electromagnetic and Yang--Mills fields. 
We are not certain about nonminimally coupled matter
since this could conceivably destroy the
very form of the junction conditions. 
Moreover in this article we stay clear of further quantum
considerations \cite{CKR} whereby both the local recovery
of SR part and of the universality of free fall of apparent-world
particles part of the equivalence principle are compromised.}
of thin branes.  This makes it a far more general condition than total 
geodesy,
since the brane apparent worlds satisfying this condition remain privileged among 
the more general
possibilities by possessing on-brane 4-d and 5-d geodesic coincidence 
and a sort of neutral
stability to falling off the brane.

\mbox{ }  We note that the absence of $Z_2$ symmetry is not immediately
disastrous in terms of matter confinement, as one may be able to
arrange the apparent world to be identified with the hypersurface
which `goes with the flow'.  However, we shall show below that this
is not generally possible. Also, even in presence of confinement,
braneworld models may suffer equivalence principle violation by the arguments above.
Finally, as we shall see below, $Z_2$ symmetry is not the only
way to ensure coincidence between geodesics and hence confinement.

One should also bear in mind that the main motivation for considering
such models is string/M-theory, within which matter species propagate
differently depending on whether they belong to open or closed string
multiplets.  This would require two distinct types of confinement.
Firstly, one would require an {\sl approximate confinement} whereby
bulk `closed string' species do not make the apparent world look
overly 5-d.  This is modelled by warping as e.g occurs in 
anti-deSitter-like bulks. Secondly, there is an independent, {\sl total
confinement} of on-brane `open string end' species.  From our 
arguments,
one way of attaining this is to invoke $Z_2$ symmetric models.  These
are neutrally stable to constituent bodies accelerating straight off
the thin matter sheet.

Despite these appealing features, there are a number of reasons why
these mechanisms may {\sl not} be satisfactory. First, the
delta function branes may come with mathematical problems,
much as delta-function particles and cosmic strings do.  Second, thick
branes \cite{thickbranes} may actually be more
desirable, \fn{Plausibility arguments for thick branes include (1)
a preference for smooth mathematics, especially as the dynamics of GR 
may not
favour the unchanging persistence of discontinuities and because smooth
mathematics may well lead to improved tractability on the long run. (2)
Quantum mechanics may not favour or be tractable for arbitrarily thin
objects.} in which case there
may still be 2 sets of privileged curves within them, without a clear
notion of which to identify with lower-d geodesics.  Thirdly, it is not
clear how string-theoretic the models considered in
this article really are.  

\mbox{ }

Testing the theories of this section presents additional difficulties.
If light is postulated to be totally confined to the 3 + 1 apparent world 
and to follow geodesics thereon, there is no difference between
light propagation on the apparent-world brane and in conventional GR.
Thus light deflection cannot be used as a distinguishing test.
For equivalence principle tests, one would need to possess a bulk matter particle in order
to compare its free fall with that of a particle made of ordinary
matter.  Were gravitons and massless scalars alone to follow 4 + 1
geodesics, massive test objects could not be built out of 4 + 1
geodesic-traversing matter so no tests of the equivalence principle could ever be
affected. In principle one could compare the paths of photons and
gravitons, but this would require observations of gravity waves and a
clear knowledge that both were emitted simultaneously.

We proceed as follows.  
The 4 + 1 bulk Einstein field equations are split up as before
with respect to a timelike hypersurface, which is moreover now declared
to be an apparent-world brane.  Thus there are additional discontinuity
terms both in the $Z_2$ case, and even more so in the asymmetric case.

We present the five `constraints' by adding and subtracting them for
the two sides of the brane.  Adding the Codazzi constraints,\fn{Here 
$j_B = (j_t, j_b) = (\mbox{\bf{T}}_{zb}, \mbox{\bf{T}}_{zt})$, i.e. 
the component of momentum flux that points into the bulk and the 3 
brane-bulk stress components. $\rho = \mbox{\bf{T}}_{zz}$, i.e. 
the bulk-bulk stress component.}   
\be
- (j_B) = D^A(K_{AB}) - D_B(K)
\mbox{ } ,
\label{C+}
\ee
and subtracting them,
\be
- [j_B] = D^A[K_{AB}] - D_B[K] = D^AY_{AB}  \mbox{ } .
\label{C-}
\ee
Adding the Gauss constraints, 
\be
R + \frac{1}{4}\left\{ (K_{AB})(K^{AB}) - (K)^2 \right\} + 2\{\rho -
\Lambda \}
= - \frac{1}{4}\left\{ [K_{AB}][K^{AB}] - [K]^2 \right\}
= - \frac{1}{4}\left\{ Y_{AB}Y^{AB} - \frac{Y^2}{3} \right\}
\mbox{ } ,
\label{G+}
\ee
and subtracting them,
\be
2[\rho - \Lambda] = (K)[K] - (K_{AB})[K^{AB}] = (K_{AB})Y^{AB}
\mbox{ } .
\label{G-}
\ee
In order to have on-brane energy-momentum conservation,
one requires $[j_B] = 0$ which gives $(j_B) = 2j_B$ in (\ref{C+}).
This precludes a net flow of energy--momentum on or 
off
the brane hypersurface. Also, as was discussed above, thick branes 
would be
prefereable in view of their smooth mathematical properties.
In that case it would be natural to have the other components of
$[\mbox{\bf{T}}_{\mbox{\scriptsize{\bf AB}}}]$ to be zero,
with $[\rho] = 0$ being of particular relevance in the above context.
Another common choice that is often made is $j_A = 0$, since
in that case the often-imposed ${Z}_2$ symmetry, $(K_{AB}) = 0$, would
allow equations (\ref{C-}) and (\ref{G-}) to be solved 
automatically
(the latter for $[\Lambda] = 0$), while at the same time simplifying
(\ref{G+}). This choice, however, entails further brane-bulk 
noninteraction.
In the case of  $j_A \neq 0$, on the other hand,
${Z}_2$ symmetry is not possible by (\ref{C+}).
In that case one would need to consider more general {\sl
asymmetric bulk} braneworld scenarios \cite{BC, asymbranes}.  

\subsection{Stability of thin matter sheets/branes}

It turns out that in addition to the matter-type-insensitive condition
$(K_{AB})= 0$, there are also matter-type-specific conditions which
ensure the coincidence of 3 + 1 and 4 + 1 geodesics.
Here we briefly consider a number of examples:

\mbox{ }

\noindent{\bf Single perfect fluid with tension}

\vskip 0.1in
\noindent We seek the conditions that a single perfect fluid
would remain on the brane, without assuming any relations
between $K^+_{AB}$ and $K^-_{AB}$. This can be obtained by
recalling that the perceived normal acceleration is the
averaged one, $<a_{z}>$, given by
\be
\{{\sigma + p}\}<a_{z}> = - \frac{1}{2}\{\lambda + p\}(K)
\mbox{ } 
\ee
where $\sigma$ and $p$ are the density and pressure of matter/energy 
of the apparent world.  
Therefore, if $\sigma \neq - p$, we obtain an expression for $<a_{z}>$,
that is zero if either the fine-tuned on-brane balance $p = -\lambda$
holds or $(K) = 0$. If $\sigma = - p$,
then either $p = -\lambda$ or $(K) = 0$ are required for consistency
and we no longer have an expression for $<a_{z}>$.
\\

A number of sub-cases can then be easily studied, including the case
of a thin sheet of dust on a hypersurface. In this case $p=0=\lambda$
and we have $<a_{z}> =0$, which shows that a thin sheet of dust stays
on whatever hypersurface it is declared to be on without any 
assumptions
about what relation there might be between $K^+_{AB}$ and $K^-_{AB}$,
as was shown by Israel \cite{Israel}.

Including pressure generalizes this condition to
\be
\{{\sigma + p}\}<a_{z}> = - \frac{p (K)}{2} \mbox{ } .
\ee
So, if $\sigma \neq - p$, we have an expression for $<a_{z}>$,
and it is only zero if $p = 0$ (the dust case) or $(K) = 0$ (the 
`stationary-average' condition).\fn{A positive-definite $K = 0$ hypersurface
is termed `maximal' (in theoretical numerical relativity, and in
closed Robertson--Walker cosmology, as in `moment of maximal expansion')
or `minimal'.  However, `stationary' is a more appropriate term 
for the indefinite version of this which occurs in this paper.  
Likewise we term indefinite $K_{AB} = 0$ hypersurfaces within spacetime 
`$z$-symmetric' in parallel with the conventional name 
`time-symmetric' for their positive-definite counterparts.} Thus even 
though Israel's result is dust-specific,
a weaker result of non-motion holds for perfect fluids, when the
two bulk extrinsic curvatures obey $(K) = 0$ which is  more general
than the `$Z_2$' or `space-symmetric average'
or `totally geodesic average' condition $(K_{AB}) = 0$.
If $\sigma = - p$ (the inflationary case), then if the universe is
nonempty, $(K) = 0$ is required for
consistency and we no longer have an expression for $<a_{z}>$.

\mbox{ }
  
\noindent {\bf Multiple perfect fluids with tension}
\vskip 0.1in
\noindent One can also consider multi-fluid models with tension, where
for example one has 
\be
Y_{AB} =  \{\sigma_1 + p_1\}u_{1A}u_{1B} - p_1g_{AB} + \{\sigma _2+
p_2\}u_{2A}u_{2B} - p_2g_{AB} - \lambda g_{AB}
\ee
for a two-component non-interacting fluid.
In this case (\ref{G+}) implies that
\be
<a_{1z}> \{ \sigma_1 + p_1 \} +
<a_{2z}> \{ \sigma_2 + p_2 \} = - \frac{p_1 + p_2 + \lambda}{2}(K)
\mbox{ } .
\label{daggers}
\ee
If neither fluid is inflationary (i.e. $\sigma_1 \neq - p_1,  \sigma_2 
\neq - p_2$), then it is impossible to ensure

\noindent
$<a_{1z}> = 0 = <a_{2z}> $ unless
\be
(K) = 0 \mbox{ or the fine-tuned on-brane balance} ~~ p_1 + p_2 =
-\lambda
\label{starry}
\ee
holds.  Furthermore, the condition $<a_{1z}> = <a_{2z}>$ is not generic, 
so the two components of the fluid will generally experience
different normal accelerations and move away
from the original brane-candidate hypersurface at different rates.

This example demonstrates that one cannot in general associate
an intially thin hypersurface of complicated enough composition with
a brane, by having it move with `the' transversal flow, as there may
simultaneously exist more than one such flow.
In the case that (without loss of generality) fluid 1 is inflationary
and fluid 2 is not, (\ref{daggers}) gives an expression for 
$<a_{2z}>$, which is zero if either of the conditions (\ref{starry}) 
hold, while we do not
have an expression for $<a_{1z}>$. If both fluids are inflationary,
then either of the conditions (\ref{starry}) must hold for consistency
and we do not have expressions for either $<a_{1z}>$ or $<a_{2z}>$.

\mbox{ }

Thus, summarizing (and in all cases assuming $p > -\sigma$),
a brane is non-accelerating if:
\vskip 0.05in
\noindent I) It is $Z_2$ symmetric, $(K_{AB}) = 0$, regardless of its
(minimally coupled) matter composition.
\vskip 0.05in
\noindent II)  It is a `stationary-average' brane, $(K) = 0$, and of
single perfect fluid composition.
\vskip 0.05in
\noindent III) It is of unrestricted $(K_{AB})$, of single perfect
fluid composition and fine-tuned in a special way.
\noindent

\mbox{ }

\noindent We also note, following \cite{BC},
that one can think of $[\rho - \Lambda]$
as an external force capable of moving around the above thin matter sheets.
We do not consider the possibility of $[\rho] \neq 0$ on the grounds of
over-richness of solutions and naturality given that 
$[j_b] = 0$ is required for on-brane energy--momentum conservation.
While this assumption changes the precise form of
the balances for no on-brane perpendicular accelerations of test
particles, it does not change the fine-tuned nature
of such balances in that they continue to require
equating on-brane matter quantities (such as $p$ and $\lambda$) with
the a priori unrelated geometrical object $(K)$ alongside the 
a priori unrelated difference of bulk cosmological constants
$[\Lambda]$.  

\mbox{ }    

\noindent{\bf Stability of branes}
\vskip 0.1in

\noindent Finally, an important question concerns the
fate of initial thin brane configurations.
The general answer to this question clearly lies beyond the
scope of the single thin brane considerations here.
Our results concerning multi-fluid cases allow some intuition
to be gained in this connection.
For example, even though pure (single component) dust
without tension would remain fixed on the initial brane,
implying some form of stability, this would not be the case for
a more general multiple component perfect fluid.
A simple example is that of 2 dust fluids with tension,
in the absence of external forces, which would usually
accelerate differently.

Further important questions in this regard are 
how thick a brane could our world be within
experimental bounds? What are the predictions of thick brane physics?
Can they thicken too quickly to agree with observations?
We note that the assumption of $Z_2$ symmetry does not
preclude thickening.  Also the neutral stability discussed above
may not be enough in general: one should also investigate 
the stability to perturbations, which may or may not be $Z_2$ symmetric,
depending upon theoretical preference.

\section{Singularities in GR and higher-d GR theories}

The physically-privileged status of geodesics makes them 
of fundamental importance in the study of singularities.  
Even though there is no universally
satisfactory definition of singularities \cite{MG,sclass},
they are often defined in terms of geodesic-incompleteness.
On the other hand, the study of singularities has been approached in terms 
of other notions (some physically motivated and some merely convenient), 
including the expansion and shear of geodesic
congruences, energy conditions and curvature scalars.

We study the cosmologies of standard GR, apparent worlds, and bulk worlds in
terms of smooth congruences of past-directed normal timelike geodesics with
normalized tangents. In each case there is a {\it normal
Raychaudhuri equation} determining the evolution of
the expansion. For example, for the standard GR we have\fn{Here, 
$\mbox{\sffamily U}^{\mbox{\sffamily{\scriptsize A}}}$,   
${\mbox{\sffamily o}}$, $\varsigma_{\mbox{\sffamily{\scriptsize AB}}}$ and 
$\mbox{\sffamily R}_{\mbox{\sffamily{\scriptsize AB}}}$
are the normal geodesic congruence flow, expansion, shear and Ricci tensor of standard GR.}
\be
\dot{\mbox{\sffamily o}} - \frac{\mbox{\sffamily o}^2}{3} =
{\varsigma}_{\mbox{\sffamily{\scriptsize
AB}}}\varsigma^{\mbox{\sffamily{\scriptsize AB}}} + \mbox{\sffamily
R}_{\mbox{\sffamily{\scriptsize AB}}}
\mbox{\sffamily U}^{\mbox{\sffamily{\scriptsize A}}}\mbox{\sffamily
U}^{\mbox{\sffamily{\scriptsize B}}} \geq 0
\mbox{ } .
\ee
Now for initial expansion $\mbox{\sffamily o}_0 > 0$ this would,
under certain global conditions, usually imply the development of 
caustics
which give rise to conjugate points (i.e the intersection points of
re-intersecting geodesics). Singularity theorems are thus obtained.
In a cosmological context the simplest of these is
\cite{Hawkingthm}\fn{See \cite{Wald} for
the technical definitions of conjugate points,
Cauchy surfaces, global hyperbolicity and the 3 energy conditions.
We note that it is the
{\it dominant energy condition} that is the most physically
motivated energy condition,
as it amounts to restricting classical
propagations to being no faster than light propagation.}

\vskip 0.1in    \noindent{\bf Cosmological Singularity Theorem 1}:
For globally-hyperbolic 3 + 1 GR spacetimes obeying the strong energy
condition 
and such that $\mbox{\sffamily o} = C \geq  0$ on some smooth
(spacelike) Cauchy surface {\sffamily S}, then no past-directed
timelike curve
from {\sffamily S} can have greater length than $\frac{3}{|C|}$.

\vskip 0.1in

\noindent By the definition of Cauchy surface, all past-directed timelike geodesics 
are among the timelike curves mentioned and are thus incomplete, so the
spacetime is singular.

There are a number of other cosmological singularity theorems
(see e.g. \cite{HE, Wald, Clarke} both to accommodate weaker assumptions
and to eliminate a number of pathological cases.
These are all phrased in terms of geodesic-incompleteness.
We recall that singularity theorems concern the existence
of singularities, saying nothing about their qualitative
properties. Carrying out a classification program of singularities
according to their properties \cite{sclass} is extremely difficult,
and may never be completed \cite{Clarke}.

Below, we confine ourselves to the commonly considered
\it curvature singularities\normalfont, for which at
least one spacetime curvature scalar such as the Ricci scalar {\sffamily R} or 
the contracted product of two Ricci tensors $\mbox{\sffamily R}_{\mbox{\sffamily{\scriptsize AB}}}\mbox{\sffamily
R}^{\mbox{\sffamily{\scriptsize AB}}}$ blows up.

Similar results also hold in higher-d. Our aim, however,
is the {\sl simultaneous} consideration of
apparent and bulk worlds, in order to understand whether a surrounding
actual bulk world can provide a different account of what appears to
be singular behaviour on the apparent world.
That some singular spacetimes may be `embeddable' in nonsingular ones
follows from simple considerations of hypersurface
geometry, which reveal that the behaviour of the higher- and lower-d
singularities need not
be related.  We consider this relation below for the simple 4 + 1 GR
models, and for braneworld models in Sec. 5.  At the outset of each 
section,
we assume that energy conditions  hold for both apparent and bulk worlds, though we 
will
consider this crucial issue toward the end of each Section.

First, as discussed in Sec. 2, the geodesics of a 3 + 1 submanifold  
are
not generally among the geodesics of a 4 + 1 embedding space.
Then the question of which curves are to be
privileged by freely falling particles translates to which curves'
extendibility is in question in the study of singularities.
A subcase of this involves 
the 3 + 1 incompleteness corresponding to the 3 + 1 apparent-world hypersurface
becoming tangent to the characteristics of the bulk equations thus
forcing the timelike and null geodesics to exit the 3 + 1 apparent-world hypersurface.
Also, the 3 + 1 geodesics might be pieces of 4 + 1 geodesics
which are extendible only by replacing a piece of the original
foliation with a new one that extends into what was originally
regarded as the extra d.  If one postulates that both
3 + 1 and 4 + 1 geodesics play a part, it is then not clear what
is meant by a singularity: exactly which curves are supposed
to be incomplete?  In particular, this plays a role in
the braneworld case of Sec. 5.

Second, the bulk expansion is\fn{Here, $\mbox{\bf h}^{\mbox{\bf{\scriptsize AB}}}$ is 
the metric induced by the bulk metric on a spatial hypersurface, 
$\mbox{\bf U}_{\mbox{\bf{\scriptsize A}}}$ is the bulk normal geodesic congruence flow and 
$U_A$ is the apparent world normal geodesic congruence flow.}
\be
\Theta  = \mbox{\bf h}^{\mbox{\bf{\scriptsize AB}}}
\Theta_{\mbox{\bf{\scriptsize AB}}}
          = \mbox{\bf h}^{zz}\mbox{\bf D}_{z}\mbox{\bf
U}_{z}
          + \mbox{\bf h}^{AB}\mbox{\bf
D}_{A}\mbox{\bf U}_{B}
          = \Theta_{zz} + h^{AB}\{D_BU_A + U_{z}K_{AB}\}
\ee
using the split provided in \cite{KucharII}.  Employing the $zz$ component of the 
definition of the bulk flow shear $\Sigma_{\mbox{\bf{\scriptsize AB}}}$, 
\be
\Theta = \Sigma_{zz} + \frac{\Theta}{4}\mbox{\bf
h}_{zz} +
\theta + U_{z}K \mbox{ } ,
\ee
where $\theta$ is the apparent-world expansion, 
so the balance between the two expansions is 
\be
\theta = \frac{3}{4}\Theta - \Sigma_{zz} - U_{z}K \mbox{ }
.
\mbox{ }
\label{expo}
\ee
This clearly demonstrates that a blow-up in ${\theta}$ need not imply
a blow-up in $\Theta$. Instead it could be the shear component 
$\Sigma_{zz}$
that blows up (Fig 1). Alternatively it could be $U_{z}$ or the trace 
$K$ of
the extrinsic curvature of the 3 + 0 slice with respect to the 4 + 0 
slice
that blow up. Thus the 4 + 1 spacetime perspective on focusing of 
geodesics
can be completely different from that of some 3 + 1 apparent-world 
hypersurface.  Thus the condition in cosmological singularity theorem 1,
$\theta \geq \frac{3}{C^{(3)}}$, neither implies nor is implied by
$\Theta \geq \frac{4}{C^{(4)}}$.
\begin{figure}[h]
\centerline{\def\epsfsize#1#2{0.6#1}\epsffile{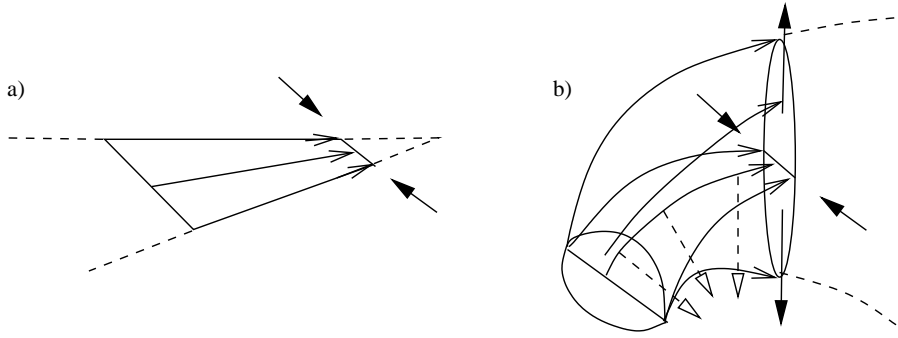}}
\caption[]{\label{TO13.ps}\footnotesize Schematic representation of how lower-d expansion can be
higher-d shear or bending.  The hollow arrows denote directed geodesics, 
the black arrows indicate how the congruence is squeezed or deformed 
and the white arrows are the normals the change in which is extrinsic curvature.\normalsize}
\end{figure}

Third, the apparent-world hypersurface components of the higher-d shear
are likewise
\be
\Sigma_{AB} = \mbox{\bf D}_{A}\mbox{\bf U}_{B} -
\frac{\Theta}{4}\mbox{\bf
h}_{AB}
              = D_AU_B + U_{z}K_{AB} - \frac{\Theta}{4}\mbox{\bf
h}_{AB}
\mbox{ } .
\ee
Using $D_AU_B = \theta_{AB} = \sigma_{AB} -
\frac{\theta}{3}h_{AB} $ together with equation (\ref{expo}) 
for the apparent-world shear $\sigma_{ab}$,
the balance of these components of the two shears is given by
\be
\sigma_{AB} = \Sigma_{AB} + \frac{1}{3}\Sigma_{zz} +
U_{z}\{
Kh_{AB} - K_{AB}\} \mbox{ } .
\ee
So a blow-up in $\sigma_{AB}$ could reflect a blow-up in $U_{z}$ or
$K_{AB}$ rather that in
$\Sigma_{\mbox{\bf{\scriptsize AB}}}$ components.

Fourth, as is already intuitively clear at the level of Gauss' 
theorem for embeddings of highly curved 2-d surfaces into 
flat 3-space,\fn{See also \cite{Shtanov} for intuition
from comparison with a curve in a plane, albeit, as stated therein,
this intuition is limited due to the curves having d = 1.} the
mathematics of embedding has a means
of `losing curvature' in transcending dimension.  To consider in
particular the balance of the bulk and apparent
Ricci scalars, we require the double contraction of
the higher-d indefinite signature version of Gauss'
theorem, namely the Gauss constraint\fn{Here, $\mbox{\bf R}_{zz}$ is a component 
of the bulk Ricci tensor, $\mbox{\bf R}$ is the bulk Ricci scalar  
and $L_{\mbox{\bf{\scriptsize AB}}}$ is the tracefree part of the extrinsic curvature 
$K_{\mbox{\bf{\scriptsize AB}}}$.}
\be
2\mbox{\bf R}_{zz} - \mbox{\bf R} = \frac{3}{4}{K}^2 -
L_{\mbox{\bf{\scriptsize AB}}}L^{\mbox{\bf{\scriptsize AB}}} - R \mbox{
}.
\label{singrem}
\ee
The 4 + 1 Einstein field equations also permit the left-hand side of this to be replaced 
by
$2\mbox{\bf T}_{zz} = 2\rho$. Thus, clearly, extrinsic curvature can
compensate for differences between higher- and lower-d intrinsic 
curvatures.
Our simple point is that one should pay attention to the implications 
for
rigorous embedding mathematics in the cases where lower-d curvature 
scalars
become infinite.  It is plausible for some 3 + 1 worlds which have 
Ricci-scalar curvature singularities $|R| \longrightarrow \infty$, to
have corresponding surrounding 4 + 1 worlds in which $\mbox{\bf R}$
(and the 5-d $\mbox{\bf T}_{zz}$) remain finite.  For this could be
compensated by $|L_{\mbox{\bf{\scriptsize AB}}}L^{\mbox{\bf{\scriptsize 
AB}}} - \frac{3}{4}{K}^2| \longrightarrow \infty$ 
which could involve
\be
|{K}| \longrightarrow \infty \left (\mbox{ i.e.~ 4 + 0 caustic 
formation }
  \right ), \mbox{and/or }
L_{\mbox{\bf{\scriptsize AB}}}L^{\mbox{\bf{\scriptsize AB}}}
\longrightarrow
\infty  \left ( \mbox{ i.e.~ shear blow up}  \right ).
\ee
If $R \longrightarrow -\infty$, shear blow up will be required.   

\subsection{Example: the embedding of flat 4-d Robertson--Walker in 5-d Minkowski spacetime}

The following simple example illustrates many of the above points and
allows further insights to be gained. Consider a 4 + 1 spacetime with 
the
metric (see e.g \cite{claims1, PDLinterp})
\be
\mbox{\bf g}_{\mbox{\bf{\scriptsize AB}}} = \mbox{diag}(\mbox{\bf
g}_{tt},
\mbox{ } h_{ab}, \mbox{ }
\mbox{\bf g}_{zz}) = \mbox{diag}
\left\{
-z^2, \mbox{ } t^{  \frac{2}{q}  }z^{  \frac{2}{1 - q}
}\delta_{ab},
\mbox{ } \frac{q^2t^2}{\{1 - q\}^2}
\right\} \mbox{ } , \mbox{ } \mbox{ } q > 1.
\label{funnymink}
\ee
This is a simple example to treat since:
\vskip 0.05in
\noindent 1) Foliating it with constant $z$ hypersurfaces,
a portion of each $z =$ const hypersurface,
has induced on it a flat Robertson--Walker metric. In
particular, with the coordinates chosen here the $z = 1$
hypersurface is the Robertson--Walker metric with scale factor $t^{\frac{1}{q}}$.
We note that $q \leq 3$ is required if the dominant energy condition is to hold.
For $q \neq 2$ there is a 3 + 1 Ricci scalar curvature singularity.
($q = 2$ is the radiation universe whence
the energy-momentum trace is zero whence R = 0.)

\vskip 0.05in
\noindent
2) The 4 + 1 spacetime is in fact Minkowski. That such an embedding
is possible has long been known \cite{Robertson} and studied elsewhere
in the literature \cite{other}.
\vskip 0.05in
To proceed we first foliate the flat Robertson--Walker metric with constant $t$ 
surfaces.
This gives $\theta = \frac{3}{q t} \longrightarrow \infty$ as
$t \longrightarrow 0$, showing that there is 3 + 1 focusing as the
Big Bang is approached.  Furthermore, {\sl only} focusing occurs:
$\sigma_{AB} = 0$. Next one foliates the 4 + 1 spacetime
with constant $t$ surfaces. Foliating the 4 + 1 spacetime
by constant-$z$ surfaces with Robertson--Walker metrics allows the
construction of a congruence by collecting the Robertson--Walker 
geodesics on each $z =$ const slice. Then the 4 + 1 expansion
${\Theta} = \frac{q + 3}{ztq}$ also blows up as
$t \longrightarrow \infty$, but there is also a blowup
of the corresponding 4 + 1 shear:
\be
\Sigma_{\mbox{\bf{\scriptsize AB}}} \equiv
\frac{q - 1}{4qzt}\mbox{diag}\left\{-h_{ab}, \mbox{ }
3\mbox{\bf
g}_{zz}\right\}
\mbox{ } .
\ee
Also, whilst both $\theta$ and $\Theta $ still blow up in this case,
the former viewed from within the $(z = 1)$ 3 + 1 
hypersurface corresponds to a genuine
3 + 1 singularity, whilst the latter is a mere caustic in 4 + 1.    

Now consider the $(z = 1)$ hypersurface as a slice of the 4 + 1 spacetime.  
The `expansion' and `shear' as regards this slicing are given by
\be
\label{caustics-shear}
K = \frac{q - 4}{q t} \mbox{ }, \mbox{ }
L_{\mbox{\bf{\scriptsize AB}}} =
- \frac{1}{4t}\mbox{diag}\left\{3, \mbox{ }
t^{\frac{2}{q}}\delta_{ab}\right\}
\mbox{ } .
\ee
Thus for the physical range of $q$, both $K$ and
$L_{\mbox{\bf{\scriptsize AB}}}$ blow up as $t \longrightarrow 0$. So 
the
spacetime includes a $(z = 0, t = 0)$ point, at which there is a 4 + 0
caustic.  These blow-ups combine in
$L_{\mbox{\bf{\scriptsize AB}}}L^{\mbox{\bf{\scriptsize AB}}}
- \frac{3}{4}{K}^2$ to cancel $R$ for all values of $q$; for $q = 2$
the `shear' and `expansion' contributions exactly cancel each other.
Note that the $q = 4$ case is an example of a pure shear blowup.
This also provides an example of how theoretically desirable candidate apparent-world
hypersurfaces in a bulk can turn out to violate energy conditions on closer inspection.  

Although the congruence discussed above is constructed to naturally
include the geodesics of all the included Robertson--Walker metrics, these turn
out not to be the 4 + 1 geodesics (nor even pieces of them). Changing
the coordinates in (\ref{funnymink}) to standard 4 + 1 Minkowski 
coordinates,
it is easy to show that the Minkowski geodesics pierce the
$z =$ constant surfaces which are the Robertson--Walker universes (more 
significantly
in the early universe).  As $t \longrightarrow 0$ the $z=$
constant surfaces approach the null cone, thus providing an example
of how foliating 3 + 1 hypersurfaces become tangent to
the characteristics at the point of interest. The foliation
breaks down as $t \longrightarrow 0$ since the family of hypersurfaces
of constant $z$ intersect at $t = 0$ [see Fig. 2].  
\begin{figure}[h]
\centerline{\def\epsfsize#1#2{0.4#1}\epsffile{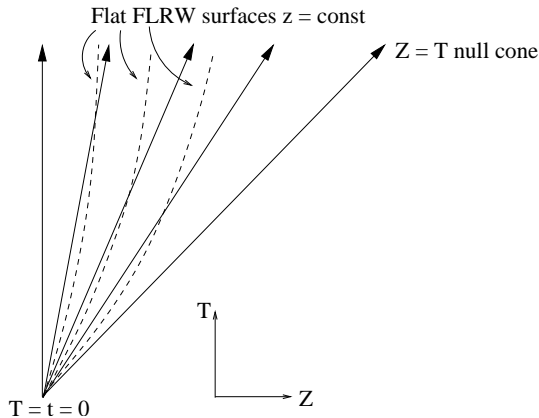}}
\caption[]{\label{TO14.ps}\footnotesize Schematic representation of the embedding of
flat Robertson--Walker universes into 4 + 1 Minkowski spacetime in the standard 
4 + 1 Minkowski coordinates $(T, X_1, X_2, X_3, Z)$.
The curved surfaces are the Robertson--Walker spacetimes. As one approaches
$T = 0$ (corresponding to $t = 0$ in Robertson--Walker coordinates), each of
these surfaces becomes tangent to the null cone (characteristics of
Minkowski).  The foliation by these surfaces also becomes bad here
because the surfaces intersect.  Note also that the Robertson--Walker geodesics
move within each of the surfaces whereas the Minkowski geodesics
clearly pierce these surfaces.  Thus the 3 + 1 and 4 + 1 geodesics
in this example are not the same.\normalsize} \end{figure}

To summarize: our studies of examples concerning the relationship 
between singularities in the 3 + 1 apparent world and 4 + 1 bulk worlds 
demonstrate that in some cases 3 + 1 singularities may be viewed as 
projective effects due to a badly-behaved choice of the foliation,
e.g one that becomes tangent to the 4 + 1 characteristics.
A clear example of this can be seen by (\ref{caustics-shear}).

In view of their special nature\fn{The examples in this paper 
all involve homogeneous and isotropic apparent worlds. The example in 
this subsection is also atypical \cite{MC} in possessing local 
embeddability into 4 + 1 Minkowski spacetime.},
however, such examples cannot be taken as representative of generic 
settings, which are of primary importance in the study of singularities.
A first potential source of general theorems are currently-known 
embedding theorems.  Unfortunately, as has been argued elsewhere 
\cite{Thanderson}, general statements based on these face  
difficulties including non-uniqueness, reliance on analytic 
functions or on high co-dimension.  They also have singularity-specific 
difficulties:  
(1) the requirement in many embedding theorems for the embedded
hypersurface to be entirely of one signature, thus excluding
the very points at which the hypersurface goes characteristic. 
(2) The impossibility of inclusion of the singularity in the set of
points on which data is prescribed, since they represent an edge
rather than a point of the spacetime. Attempts at their inclusion
by providing data `right up' to the edge, cannot always stop the
data from becoming badly-behaved, by e.g some components of
$K_{{\mbox{{\scriptsize AB}}}}$ becoming infinite when attempting
to embed a Ricci scalar singular spacetime into a nonsingular 
one.\fn{These intuitions generalize those in \cite{Shtanov}; a 
valuable account of some of these issues is also presented in 
\cite{claims2}} 
(3) The approach to singularities is not always analytic \cite{Clarke}.

A second potential source of general theorems are the classical 
singularity theorems.  These readily generalize to higher-d 
spacetimes, provided that analogous energy conditions  are obeyed.
However, then even if one were to succeed in excising
singularities from a lower-d model, one would typically expect
singularities to occur {\sl elsewhere} in the resultant higher-d models
by the higher-d singularity theorems.  Also the nature of the 
corresponding higher-d singularities could be different, specially 
given that the unrestricted embedding of symmetric spacetimes may 
result in `less symmetric' higher-d models which by analogy with 
\cite{Kras}  are capable of exhibiting
a wider range of singular (and even nonsingular)
behaviours  \cite{Kras}.  We add that from the GR perspective 
it may be viewed as unsatisfactory to replace the unambiguous 
geodesics with some new privileged curves whose nature depends 
on the highly nonunique procedure that can be involved in embeddings.

Finally, a key question concerns the nature of bulk matter
and the corresponding energy conditions, specially given the absence of 
observational or theoretical constarints at present.  
This is important since an assumed bulk strong energy
condition need not induce an 
apparent-world strong energy
condition  nor vice versa. Minkowski bulks in which anti-deSitter apparent worlds 
are embedded suffice to illustrate the former, and anti-deSitter bulks in 
which Minkowski apparent worlds are embedded suffices to illustrate the latter.
The crucial point is that the bulk expression
$\mbox{\bf R}_{\mbox{\bf{\scriptsize AB}}}
\mbox{\bf U}^{\mbox{\bf{\scriptsize A}}}
\mbox{\bf U}^{\mbox{\bf{\scriptsize B}}}$
contains not only the apparent-world expression $R_{AB}U^AU^B$ but also a 
complicated - albeit easily-computible - 
combination of terms (involving $\mbox{\bf R}_{Az}$, 
$\mbox{\bf R}_{zz}$, $K_{AB}$ etc) that arise from the projection of 
the 4 + 1 Ricci tensor.  A related point concerning the strong energy
condition is that from the stability condition for geodesics near a 
totally-geodesic surface \cite{Seahra},
\be
\mbox{\bf R}_{\mbox{\scriptsize{\bf AB}}}{\mbox{\bf 
U}}^{\mbox{\scriptsize{\bf A}}}{\mbox{\bf U}}^{\mbox{\scriptsize{\bf B}}}
> R_{AB}U^{A}U^{B} \mbox{ for stability }
\ee
and the choice of bulk field equations   
$\mbox{\bf R}_{\mbox{\scriptsize{\bf AB}}} = 0$,
one cannot have totally geodesic submanifolds that are simultaneously
stable in the above sense and obeying the strong energy
condition. Moreover, if there are 
two
distinct notions of privileged curves, the ordinary notion of
causality, which physically motivates the dominant energy condition, is undermined.

Note that these complications involving lower- and higher-d energy conditions
{\sl compromise the apparent-world versions of, among other theorems, the classical
cosmological singularity theorems \cite{Wald}.}  
The points discussed in the last three paragraphs 
furthermore apply just as well to the case of
braneworlds, to which we now turn.

\section{Singularities in braneworld models}

The fact that 3 + 1 geodesics do not generally coincide with those in 4 
+ 1
complicates the study and characterization of
singularities in these settings.  In the case of models satisfying
$Z_2$ symmetry the geodesics {\sl do} coincide, thus removing many
of these complications. This provides an important motivation
for the consideration of $Z_2$ symmetric models.

Let the normal congruences of the bulk and apparent world have normalized tangents denoted by 
\be
\mbox{\bf U}_{\mbox{{\scriptsize\bf A}}}^{\mbox{\scriptsize bulk}} =
\mbox{\bf U}_{\mbox{{\scriptsize\bf A}}}^{\mbox{\scriptsize bulk
species}} +
\mbox{\bf U}_{\mbox{{\scriptsize\bf A}}}^{\mbox{\scriptsize brane
species}}\delta^{(4)}(z) 
\mbox{ } \mbox{ } , \mbox{ } \mbox{ }
U_{A}^{\mbox{\scriptsize brane}} =
U_{A}^{\mbox{\scriptsize bulk species}} +
U_{A}^{\mbox{\scriptsize brane species}}
\mbox{ } .
\ee
The balance of expansions $\theta$ and $\Theta$ can then be arranged to 
give
\bea
\frac{3}{4}(\Theta) - (\Sigma_{zz}) +
U_{z}^{\mbox{\scriptsize bulk
species}}(K) &=&
2\theta \mbox{ } , \\  \mbox{ }
\frac{3}{4}[\Theta] = [\Sigma_{zz}] +
U_{z}^{\mbox{\scriptsize bulk
species}}\frac{Y}{6} \mbox{ }  ,
\eea
using the fact that $\mbox{\bf U}_{z}^{\mbox{\scriptsize brane 
species}}$
does not contribute as well as the continuity requirement

\noindent
$[\mbox{\bf U}_{z}^{\mbox{\scriptsize bulk species}}] =
0$.  In the $Z_2$ case with $\mbox{\bf U}_{z}^{\mbox{\scriptsize
bulk-traversing species}} = 0$, there is less freedom to compensate
for on-brane expansion blow-ups than in the previous section:
\bea
\frac{3}{4}(\Theta) - (\Sigma_{zz}) &=&
2\theta
\mbox{ } ,
\\
\mbox{ }
\frac{3}{4}[\Theta] = [\Sigma_{zz}] 
\mbox{ } .
\eea
Freedom to compensate is also restricted if the brane species have no
isotropic contribution $Y = 0$ or if the brane is stationary on average, 
$(K) = 0$.  

The balance of apparent-world hypersurface components of the shears are given by
\be
2\sigma_{AB} = (\Sigma_{AB}) + \frac{(\Sigma_{zz})}{3}h_{AB} +
U_{z}^{\mbox{\scriptsize bulk species}}\{(K)h_{AB} - (K_{AB})\}
\ee
\be
0 = [\Sigma_{AB}] + \frac{(\Sigma_{zz})}{3}h_{AB} +
U_{z}^{\mbox{\scriptsize bulk species}}\frac{Y_{AB}}{2} \mbox{ } .
\ee
So $Z_2$ symmetry, $\mbox{\bf U}_{z}^{\mbox{\scriptsize bulk species}} 
= 0$,
and no brane species matter $Y_{AB} = 0$ are conditions which restrict
freedom to compensate for on-brane shear blow-ups.

As regards the prospect of Ricci scalar curvature singularity removal,
the balance equation in the $Z_2$ symmetric case is now given by
\be
\frac{1}{4} \left\{ \frac{Y^2}{3} - Y_{AB}Y^{AB} \right\} - R +
2\Lambda
     =   2\mbox{\bf R}_{zz} - \mbox{\bf R}
     =   2\rho
\mbox{ } .
\label{number}
\ee
Now if $|R| \longrightarrow \infty$ and $\mbox{\bf R}$ and $\rho$ are
to remain finite, then we require
$\left|\frac{Y^2}{3} - Y_{AB}Y^{AB}\right| \longrightarrow \infty$.
It is interesting to consider under which circumstances the
brane-confined fields' energy-momentum tensor $Y_{AB}$ remains finite
or becomes infinite. It is the former which is unusual as
far as standard cosmology is concerned.
Although compensation for Ricci scalar blow-up is still possible, while
$K_{AB}$ remains finite, through the inverse metric blowing up,
$Y_{AB} =-K_{AB} + Kh_{AB}$ will nevertheless diverge if the
blow-up is a caustic $|K| \longrightarrow \infty$.
Finite density of
brane-confined fields is only possible if the blow-up is pure `shear'
$|L_{AB}L^{AB}| \longrightarrow \infty$ and due to $h^{AB}$ rather than
$L_{AB}$ blowing up.

Also, quite unlike GR, there is a possibility here for two quantities,
which are measurable from within the apparent world, to both blow up and yet
produce a bulk world which is nonsingular, without resorting to
assumptions about any inaccessible bulk parameters.
Thus given the novel physical interpretation placed on $[K_{AB}]$
in braneworld models, the Ricci scalar $R$
and brane-confined fields' energy-momentum $Y_{AB}$ could both blow up,
but in such a way that the first and second terms on the left hand side
of (\ref{number}) cancel each other.

In the non-$Z_2$ case, there is an extra term $(K_{AB})(K^{AB}) - 
(K)^2$
on the left hand side which could be used to balance the
Ricci scalar singularity. The brane-confined
species' energy-momentum tensor $Y_{AB}$ is free to remain
finite in this case, as it is unrelated to $(K_{AB})$.  

\subsection{Examples}

The above considerations allow a clearer understanding
of specific examples of `unusual' singular behaviour considered in
the literature. One recent example is given in \cite{Shtanov}, where
in the light of our discussions above it is the
curvature blow-ups that are compensated by the extrinsic
curvature blow-ups, thus keeping 4 + 1 Ricci curvature finite.
We can also answer the question raised in that paper, namely whether
braneworlds that are more general than isotropic can have unusual
singularities. According to our results here, the answer is in 
principle
yes, in the sense that the required balances are in principle possible
for the fully general braneworld field equations in the $Z_2$ case and even more so
in the non-$Z_2$ case. What is not, however, known a priori is whether 
these
possibilities {\sl typically occur} for the fully general braneworld 
field equations.

As another example, we further study the example of a bulk anti-deSitter 
metric given in \cite{Ishihara}:
\be
\mbox{\bf g}_{\mbox{\bf{\scriptsize AB}}} =
\mbox{diag}(\mbox{\bf
g}_{tt},
\mbox{ } h_{ab}, \mbox{ }
\mbox{\bf g}_{zz}) =
\mbox{diag}\left\{ - \left\{ 1 + \frac{z^2}{l^2} \right\},
z^{2}S_{ab}^{(3)}, \left\{ 1 +
\frac{z^2}{l^2} \right\}^{-1} \right\}
\label{ads}
\ee
(with coordinates $t$, $\theta$, $\phi$, $\psi$, $z$),
where $l$ is some `anti-deSitter length' closely related to the bulk cosmological
constant $\Lambda$ and ${S_{ab}}^{(3)}$ is the 3-sphere which we
coordinatize by the angles $\theta$, $\phi$, $\psi$.
This bulk contains a brane that is a closed Robertson--Walker universe
obtained by setting $r = a$ to be the
solution of the braneworld version of the Friedmann equation and $t$ to
be the solution of $- (1 + a^2/l^2) (dt/d\tau)^2 +
(1 + a^2/l^2)(da/d\tau)^2 = -1$.

Using $a = t$ and leading on from a comment in \cite{Ishihara},
our first point is that this means that the brane becomes tangent to
the characteristics of the bulk anti-deSitter spacetime at these singularities.
Consequently certain embedding theorems are not applicable there.

Secondly, using the asymptotic solution given in \cite{Ishihara},
we find that the asymptotic form of the induced 4-metric,
\be
h_{AB} = \mbox{diag}\left\{ - \left\{ 1 + \frac{t^2}{l^2} \right\},
t^{2}S^{(3)}_{ab} \right\}
\label{adsbrane}
\ee
(with coordinates $t$, $\theta$, $\phi$, $\psi$),
obtained by using $z = a = t$ in (\ref{ads}), has a Ricci curvature
singularity $R \sim 1/t^2$ as $t \longrightarrow 0$
which corresponds in each case (each with its own choice of $t$) to the
approach to the singularity.  On the other hand,
the anti-deSitter bulk remains perfectly finite there.  What happens is that both
$L_{\mbox{{\scriptsize AB}}}L^{\mbox{{\scriptsize AB}}}$ and $K^2$ blow
up as $\sim 1/t^2$ so as to compensate for the on-brane Ricci
curvature singularity. Moreover, in
this example, $K_{AB}$ remains finite, while it
is the inverse induced metric that blows up.  Nevertheless, the
brane-confined matter energy-momentum tensor
blows up, because $Y_{AB} = - K_{AB} + Kh_{AB}$ and $K$ blows up.
This provides an example of a singularity from the on-brane
perspective which is balanced from the bulk
perspective such that the blow-up of the brane-confined fields'
energy-momentum cancels out the blow-up of the
on-brane Ricci scalar.

Finally, other on-brane Robertson--Walker universes have been considered
\cite{Shtanov,DD00} where derivatives of the scale factor blow-up
rather than the scale factor shrinking to zero. These serve as 
additional
examples of behaviours qualitatively different from the Big Bang, 
but are not connected to the types of balances discussed in this paper.

To summarize, we have shown that there is definitely scope in 
braneworld models
for two distinct notions of physically privileged curves to coexist.
Thus the notion of `singular' is even more complicated than in GR case
since one would have to consider the extendibility for two
congruences of privileged curves of different dimension.   
Also, as pointed out in \cite{Caldwelletc, Ishihara, CKR}, assigning
roles to both higher-d and lower-d
geodesics leads in some cases to amelioration of the apparent world's
horizon problem. We should point out that this too is not 
solely a feature of braneworlds, but that it is
shared by the simpler, not totally-geodesic,
models considered in the previous section.
We also point out that whenever this amelioration
occurs, then the apparent-world dominant energy condition loses physical motivation.  
Conversely, when some 3 + 1
geodesics permit faster travel than those in 4 + 1, the bulk dominant energy condition loses 
physical 
motivation. This is a disturbing legacy given
the technical importance of energy conditions, and in particular the dominant energy condition
which is currently required, for example, for the
positive energy theorem and the zeroth law of black hole mechanics to
hold.  So such a resolution of the horizon problem would come
hand-in-hand with losing understanding of compact object astrophysics.

We also emphasize that the strong energy
condition is compromised in braneworlds, both as
in Sec. 4.1 as well as for other braneworld reasons. To see this
consider the apparent-world Raychaudhuri equation
\be
\dot{\theta} = - \frac{\theta^2}{3} +
\left\{(K)(K_{AB}) - (K_{AC})({K^C}_B)
+
\mbox{Lin}_{AB} + \mbox{Quad}_{AB}
\right\}
U^AU^B
\ee
where Lin$_{AB}$ and Quad$_{AB}$ are linear and quadratic matter terms
respectively. Thus even in the $Z_2$ case the presence of
quadratic matter terms can interfere with establishing
an inequality along the lines of the strong energy
condition, while in the
non $Z_2$ case geometrical terms that are untied to matter content can
additionally interfere.
Thus in GR-based braneworlds, apparent-world versions of the classical
singularity theorems (listed in \cite{Wald}) appear
compromised.\fn{The issue of strong energy
condition violation in braneworlds was also
raised in \cite{Muk}.}
One problem that can be alleviated in braneworlds is that for bulks
with a pure negative cosmological constant,
our conclusion in Sec. 4.2, using the stability results from
\cite{Seahra} concerning the stability of
geodesics on a submanifold to bulk perturbations
not being compatible with the strong energy
condition on that manifold, does not hold.  
This is another reason to choose negative cosmological
constant bulks.

Moreover, other complications can occur for certain braneworld
models with anti-deSitter-like bulks, which though desirable for their confining
properties, do not satisfy the weak energy condition.
This compromises the bulk versions of one of the classical
singularity theorems
  and of the third law of black hole mechanics.
Furthermore, anti-deSitter-like bulks have Cauchy horizons \cite{horiz} and thus
are not globally hyperbolic. This compromises the bulk version of cosmological singularity theorem 1,
although the other classical singularity theorems (see \cite{Wald})  
survive this pitfall since they do not require global hyperbolicity.
It should also be noted that `global hyperbolicity of
the apparent world' may be compromised through bulk influences.   
This would cause an apparent-world analogue of cosmological singularity theorem 1 
to lose its motivation.

\section{Conclusion}

We have shown that the geodesics of a codimension 1 submanifold are not
generically among those of the manifold.  
One consequence is that the contents of a given submanifold which has been 
declared as the apparent world may have a propensity to fall off it.
Another is that the equivalence principle may no longer hold if 3 + 1
GR is replaced by such a submanifold embedded in a higher-d bulk.
For, free fall may acquire an element of composition dependence or
may cease to be along the 3 + 1 timelike geodesics of the submanifold.  
Even though it is true that totally-geodesic submanifolds have geodesics 
which coincide with those of the surrounding bulk, such worlds are not generic.
Also it is not clear that submanifolds that are not branes can serve as 
apparent worlds.  
In principle observers may move on a submanifold by appropriate
choice of motion. 
The question, however, is what {\sl obliges} observersto move on that submanifold? 
Can matter leave that submanifold even if undergoing forced (rather than geodesic) motion 
and would an observer whose motion is restricted to that submanifold (by choice or otherwise) 
notice its disappearance?

In models with branes, however, there is confinement by fiat (some 
species are declared to be brane-confined fields and observers could be confined by
virtue of being made out of these - e.g Standard Model fields) while
there is also approximate confinement by warping [whereby
bulk-traversing species look very (3 + 1)-d also].
There is evidence that this approximate confinement
is within acceptable experimental bounds for at least some braneworld
models \cite{branes, GarrTan}. 
For a general thin asymmetric braneworld,
bulk and brane geodesics will not
in general coincide, so bulk-propagating and brane-confined matter
test-bodies undergo distinct free fall: there is in principle equivalence principle
violation and on-brane gain or loss of bulk species matter.
Moreover, at the level of these GR-like braneworld models,
$Z_2$-symmetric braneworlds have apparent and bulk geodesic
coincidence on the apparent-world brane while being considerably more general
that totally-geodesic submanifolds.  
Thus we argue that these are `more like GR'
than asymmetric brane models, which makes them more plausible
in themselves within the GR-like braneworld setting, quite apart from
$Z_2$ symmetry being a feature of at least some phenomenological
string-theoretic scenaria \cite{HW}.  
Establishing that was the
first aim of this article, alongside constructing a list of
aspects of asymmetric braneworlds that are
rather more complicated than one might na\"{\i}vely expect from
studying $Z_2$ worlds.  
We also point out that it is not entirely
clear whether the idea of brane-confined matter is compatible with GR,
in the sense that were the framework of study general enough to include
thick branes, could initially-thin branes rapidly become thick?

As regards singularities, we showed that in brane-free models there are already
simple balances whereby the bulk and apparent-world versions of the objects used
in the study of singularities may greatly differin behaviour.  
This affects curvature singularities and, more
significantly, the energy conditions  on which many theorems rest.  
While these balances permit some curvature singularities to be removed
by transcending dimension, this can sometimes be mere alteration or 
displacement rather than true removal of singularities.  
We showed that all these phenomena are compounded by including effects due to branes.  
Again, non-$Z_2$ braneworlds open up new possibilities while complicating the
definition of a singularity, by complicating the issue of exactly which
curves are supposed to be incomplete.  
On the other hand $Z_2$ braneworlds are more reassuringly GR-like.
Nevertheless, even these are subject to difficulties
due to the bulk and apparent energy conditions not matching up.

\mbox{ }  

\noindent{\bf Acknowledgements}
\vskip 0.1in

\noindent We would like to thank
Malcolm MacCallum, Gary Gibbons and
Sanjeev Seahra for fruitful discussions, and
Clifford Will and Jurgen Ehlers for comments.
RT thanks Miguel Vazquez-Mozo and Kerstin Kunze for helpful discussions.
EA thanks PPARC, Peterhouse Cambridge and the Killam Foundation for
funding over the years in which this work was done, and Queen
Mary London for Visitor status.

\mbox{ }

\end{document}